\documentclass{aastex63}

\newcommand{\ie}{{\it i.e.}}
\newcommand{\eg}{{\it e.g.}}
\newcommand{\etal}{{\it et al.}}

\newcommand{\kepler}{{\it Kepler}}
\newcommand{\Kepler}{{\it Kepler}}

\newcommand{\Ktwo}{{\it K2}}
\newcommand{\ktwo}{\Ktwo}

\newcommand{\tess}{{\it TESS}}

\newcommand{\Gaia}{{\it Gaia}}
\newcommand{\gaia}{{\it Gaia}}

\newcommand{\teff}{$T_{\mathrm{eff}}$}

\newcommand{\prot}{$P_{\mathrm{rot}}$}

\newcommand{\degrees}{$^\circ$}
\newcommand{\vz}{$v_{\bf z}$}
\newcommand{\vb}{$v_{\bf b}$}
\newcommand{\kms}{kms$^{-1}$}
\newcommand{\sigmavb}{$\sigma_{v{\bf b}}$}
\newcommand{\sigmavz}{$\sigma_{v{\bf z}}$}

\newcommand{\gcolor}{$G_{BP} - G_{RP}$}

\newcommand{\mct}{\citet{mcquillan2014}}

\newcommand{\racomment}[1]{{\color{black}#1}}

\begin{document}

\title{Exploring the evolution of stellar rotation using Galactic kinematics}


\correspondingauthor{Ruth Angus}
\email{rangus@amnh.org}

\author[0000-0003-4540-5661]{Ruth Angus}
\affiliation{Department of Astrophysics, American Museum of Natural History,
200 Central Park West, Manhattan, NY, USA}
\affiliation{Center for Computational Astrophysics, Flatiron Institute,
162 5th Avenue, Manhattan, NY, USA}
\affiliation{Department of Astronomy, Columbia University, Manhattan, NY, USA}

\author[0000-0002-8658-1453]{Angus~Beane}
\affiliation{Center for Astrophysics $\vert$ Harvard \&
Smithsonian, 60 Garden Street, Cambridge, MA 02138, USA}

\author[0000-0003-0872-7098]{Adrian~M.~Price-Whelan}
\affiliation{Center for Computational Astrophysics, Flatiron Institute,
162 5th Avenue, Manhattan, NY, USA}

\author[0000-0003-4150-841X]{Elisabeth Newton}
\affiliation{Department of Physics and Astronomy, Dartmouth
College, Hanover, NH, USA}

\author[0000-0002-2792-134X]{Jason L. Curtis}
\affiliation{Department of Astrophysics, American Museum of Natural History,
200 Central Park West, Manhattan, NY, USA}

\author[0000-0002-2580-3614]{Travis Berger}
\affiliation{Institute for Astronomy, University of Hawai`i, 2680
Woodlawn Drive, Honolulu, HI 96822, USA}

\author[0000-0002-4284-8638]{Jennifer van Saders}
\affiliation{Institute for Astronomy, University of Hawai`i, 2680
Woodlawn Drive, Honolulu, HI 96822, USA}

\author[0000-0003-2102-3159]{Rocio Kiman}
\affiliation{Department of physics, CUNY Graduate Center, City
University of New York, Manhattan, NY, USA}
\affiliation{Department of Astrophysics, American Museum of Natural History,
200 Central Park West, Manhattan, NY, USA}

\author[0000-0002-9328-5652]{Daniel Foreman-Mackey}
\affiliation{Center for Computational Astrophysics, Flatiron Institute,
162 5th Avenue, Manhattan, NY, USA}

\author[0000-0003-4769-3273]{Yuxi (Lucy) Lu}
\affiliation{Department of Astronomy, Columbia University, Manhattan, NY, USA}
\affiliation{Department of Astrophysics, American Museum of Natural History,
200 Central Park West, Manhattan, NY, USA}

\author[0000-0001-5725-9329]{Lauren Anderson}
\affiliation{Observatories of the Carnegie Institution for
Science, 813 Santa Barbara Street, Pasadena, CA 91101, USA}

\author[0000-0001-6251-0573]{Jacqueline K. Faherty}
\affiliation{Department of Astrophysics, American Museum of Natural History,
200 Central Park West, Manhattan, NY, USA}

\begin{abstract}
The rotational evolution of cool dwarfs is poorly constrained after $\sim$1-2
Gyr due to a lack of precise ages and rotation periods for old main-sequence
stars.
\racomment{In this work we use velocity dispersion as an age proxy to reveal the
temperature-dependent rotational evolution of low-mass \kepler\ dwarfs, and
demonstrate that kinematic ages could be a useful tool for calibrating
gyrochronology in the future.}
We find that a linear gyrochronology model, calibrated to fit the
period--\teff\ relationship of the Praesepe cluster, does not apply to stars
older than around 1 Gyr.
Although late-K dwarfs spin more slowly than early-K dwarfs when they are
    young, at old ages we find that late-K dwarfs rotate at the {\it same
    rate} or faster than early-K dwarfs of the same age.
This result agrees qualitatively with semi-empirical models that vary the rate
of surface-to-core angular momentum transport as a function of time and mass.
It also aligns with recent observations of stars in the NGC 6811 cluster,
which indicate that the surface rotation rates of K dwarfs go through an epoch
of inhibited evolution.
We find that the oldest \kepler\ stars with measured rotation periods are
late-K and early-M dwarfs, indicating that these stars maintain spotted
surfaces and stay magnetically active longer than more massive stars.
Finally, based on their kinematics, we confirm that many rapidly rotating
GKM dwarfs are likely to be synchronized binaries.
\end{abstract}


\section{Introduction}

\subsection{Gyrochronology}

Stars with significant convective envelopes ($\lesssim$ 1.3 M$_\odot$) have
strong magnetic fields and slowly lose angular momentum via magnetic braking
\citep[\eg][]{schatzman1962, weber1967, kraft1967, skumanich1972, kawaler1988,
pinsonneault1989}.
Although stars are born with random rotation periods, between 1 and 10 days,
observations of young open clusters reveal that their rotation periods
converge onto a unique sequence by $\sim$500-700 million years
\citep[\eg][]{irwin2009, gallet2013}.
After this time, the rotation period of a star is thought to be determined, to
first order, by its color and age alone.
This is the principle behind gyrochronology, the method of inferring a star’s
age from its rotation period \citep[\eg][]{skumanich1972, barnes2003,
barnes2007, mamajek2008, barnes2010, meibom2011, meibom2015}.
However, new photometric rotation periods made available by the \kepler\
\citep{borucki2010} and \ktwo\ \citep{howell2014} missions
\citep[\eg][]{mcquillan2014, garcia2014, douglas2017, rebull2017, meibom2011,
meibom2015, curtis2019} confirm that rotational evolution is a highly complex
process.
For example, the early-to-mid M dwarfs in the $\sim$ 650 Myr Praesepe cluster
spin more slowly than the G dwarfs; in theory because lower-mass stars have
deeper convective zones which generate stronger magnetic fields and more
efficient magnetic braking.
However, in the NGC 6811 cluster which is around 1 Gyr \citep{janes2011,
sandquist2016},
late-K dwarfs rotate at the {\it same rate} as early-K dwarfs
\citep{curtis2019}.
\racomment{In addition, new rotation period measurements for low-mass stars in
the 2.7-Gyr-old cluster Ruprecht 147 show that its slow-rotator sequence is
flat compared to younger clusters like Praesepe (Curtis \etal\ 2020, submitted
to AAS Journals), and departures from classical Skumanich-like spin-down were
also noted in observations of the $\sim$1.3 Gyr NGC 752 open cluster
\citep{agueros2018}.}
New semi-empirical models that vary the rate of angular momentum
redistribution in the interiors of stars are able to reproduce the flattened
period--color relation \racomment{seen in these clusters \citep{spada2019}.}
These models suggest that mass and age-dependent angular momentum transport
between the cores and envelopes of stars has a significant impact on their
surface rotation rates.

Another example of unexpected rotational evolution is seen in old field stars
which appear to rotate more rapidly than classical gyrochronology models
predict \citep{angus2015, vansaders2016, vansaders2018, metcalfe2019}.
A mass-dependent modification to the classical \prot\ $\propto
t^{\frac{1}{2}}$ spin-down law \citep{skumanich1972} is required to reproduce
these observations.
To fit magnetic braking models to these data, a cessation of magnetic braking
is required after stars reach a Rossby number (Ro; the ratio of rotation
period to convective turnover time) of around 2 \citep{vansaders2016,
vansaders2018}.

The rotational evolution of stars is clearly a complicated process and, to
fully calibrate the gyrochronology relations we need a large sample of
reliable ages for stars spanning a range of ages and masses.
In this paper, we use the velocity dispersions of field stars to qualitatively
explore the rotational evolution of GKM dwarfs, and show that kinematics could
provide a gyrochronology calibration sample.

\subsection{Using velocity dispersion as an age proxy}

Stars are thought to be born in the thin disk of the Milky Way (MW), orbiting
the Galaxy with a low out-of-plane, or vertical, velocity (\vz),
just like the star-forming molecular gas observed in the disk today
\citep[\eg][]{stark1989, stark2005, aumer2009, martig2014, aumer2016}.
On average, the vertical velocities of older stars is observed to be larger
\citep[\eg][]{nordstrom2004, holmberg2007, holmberg2009, aumer2009,
casagrande2011}.
This is likely either a signature of dynamical heating, such as from
interactions with giant molecular clouds, spiral arms and the galactic bar
\citep[see][for a review of secular evolution in the MW]{sellwood2014}, or an
indication that stars formed dynamically ``hotter'' in the past
\citep[e.g.,][]{bird2013}.
In either case, the vertical velocity distribution is observed to depend
significantly on stellar age.
While the velocity of any individual star only provides a weak age constraint
\racomment{(if any at all)}, because its velocity depends on its current
position in its orbit, the velocity {\it dispersion} of a {\it population} of
stars indicates whether that population is old or young relative to other
populations.
In this work, we compare the velocity dispersions of populations of field
stars in the Galactic thin disk to ascertain which populations are older and
which younger, and draw conclusions about the rotational evolution of stars
based on their implied relative ages.

\racomment{There is a long history of using kinematic ages to explore the
evolution of cool dwarfs \citep[\eg][]{reid1995, gizis2000, west2004,
west2006, schmidt2007, faherty2009, kiman2019}.
For example, \citet{west2004, west2006} used the vertical distances of stars
from the Galactic mid-plane as an age proxy, and found that the fraction of
magnetically active M dwarfs decreases over time.
\citet{faherty2009} used tangential velocities to infer the ages of M, L and T
dwarfs, and showed that dwarfs with lower surface gravities tended to be
kinematically younger, and \citet{kiman2019} used velocity dispersion as an
age proxy to explore the evolution of H$\alpha$ equivalent width (a magnetic
activity indicator), in M dwarfs.
}

Although {\it vertical} velocity, \vz, is \racomment{a well-established} age
proxy, it can only be calculated with full 6-dimensional position and velocity
information.
In fact, with full 6D phase space and an assumed Galactic potential, it is
possible to calculate the dynamically-invariant vertical {\it action}, which may
be an even better age indicator \citep{beane2018, ting2019}.
Unfortunately, most field stars with measured rotation periods do not have
radial velocity (RV) measurements because they are relatively faint \kepler\
targets ($\sim$12th-16th magnitudes).
For this reason, we used velocity in the direction of galactic latitude, \vb,
as a proxy for \vz.
The \kepler\ field is positioned at low galactic latitude
(b=$\sim$5-20\degrees), so \vb\ is a close (although imperfect, see
appendix) approximation to \vz.
Because we use \vb\ rather than \vz\, we do not calculate absolute kinematic
ages using a published age--velocity dispersion relation (AVR), calibrated with
vertical velocity.
In the future it may be possible to account for the differences between \vb\
and \vz, or marginalize over missing RV measurements and the \kepler\
selection function, in order to infer the absolute ages of populations of
stars.
Regardless of direction however, velocity dispersion is expected to
monotonically increase over time \citep[\eg][]{holmberg2009}, and can
therefore be used to {\it rank} populations of stars by age.

This paper is laid out as follows: in section \ref{sec:method} we describe our
sample selection process and the methods used to calculate stellar
velocities.
In section \ref{sec:results} we use kinematics to investigate the relationship
between stellar rotation period, age and color/\teff\ and interpret the
results.
We also examine the rotation period gap and the kinematics of synchronized
binaries.
In the appendix, we establish that \vb\ velocity dispersion, \sigmavb, can be
used as an age proxy by demonstrating that neither mass-dependent heating nor
the \kepler/\gaia\ selection function is observed to strongly affect our
sample.
The data used in this project is described in table \ref{tab:data}, at the end
of this paper, and is available online.

\section{Method}
\label{sec:method}

\subsection{The data}
\label{sec:the_data}

We used the publicly available \kepler-\gaia\ DR2 crossmatched
catalog\footnote{Available at gaia-kepler.fun} to combine the \mct\ catalog of
stellar rotation periods, measured from \kepler\ light curves, with the \gaia\
DR2 catalog of parallaxes, proper motions and apparent magnitudes.
Reddening and extinction from dust was calculated for each star using the
Bayestar dust map implemented in the {\tt dustmaps} {\it Python} package
\citep{green2018}, and {\tt astropy} \citep{astropy2013, astropy2018}.
For this work, we used the precise \textit{Gaia} DR2 photometric color,
$G_{\rm BP} - G_{\rm RP}$, to estimate \teff\ for the Kepler rotators.
\racomment{The calibration of this relation is described in Curtis \etal\
(2020, in prep) and summarized in the Appendix of this paper.}

\begin{figure}
  \caption{
      Top: de-reddened MS \kepler\ stars with \mct\ rotation periods, plotted
    on a \gaia\ CMD.
    We removed photometric binaries and subgiants from the sample by excluding
    stars above the dashed lines.
    Bottom: a zoom-in of the top panel, with stars colored by their
    gyrochronal age \citep{angus2019}, instead of their rotation period.
    A general age gradient is visible across the main sequence.
    Since the \citet{angus2019} relation predicts that the oldest stars in
    the \mct\ sample are late-G and early-K dwarfs, it is probably
    under-predicting the ages of late-K and early-M dwarfs.
}
  \centering
    \includegraphics[width=1\textwidth]{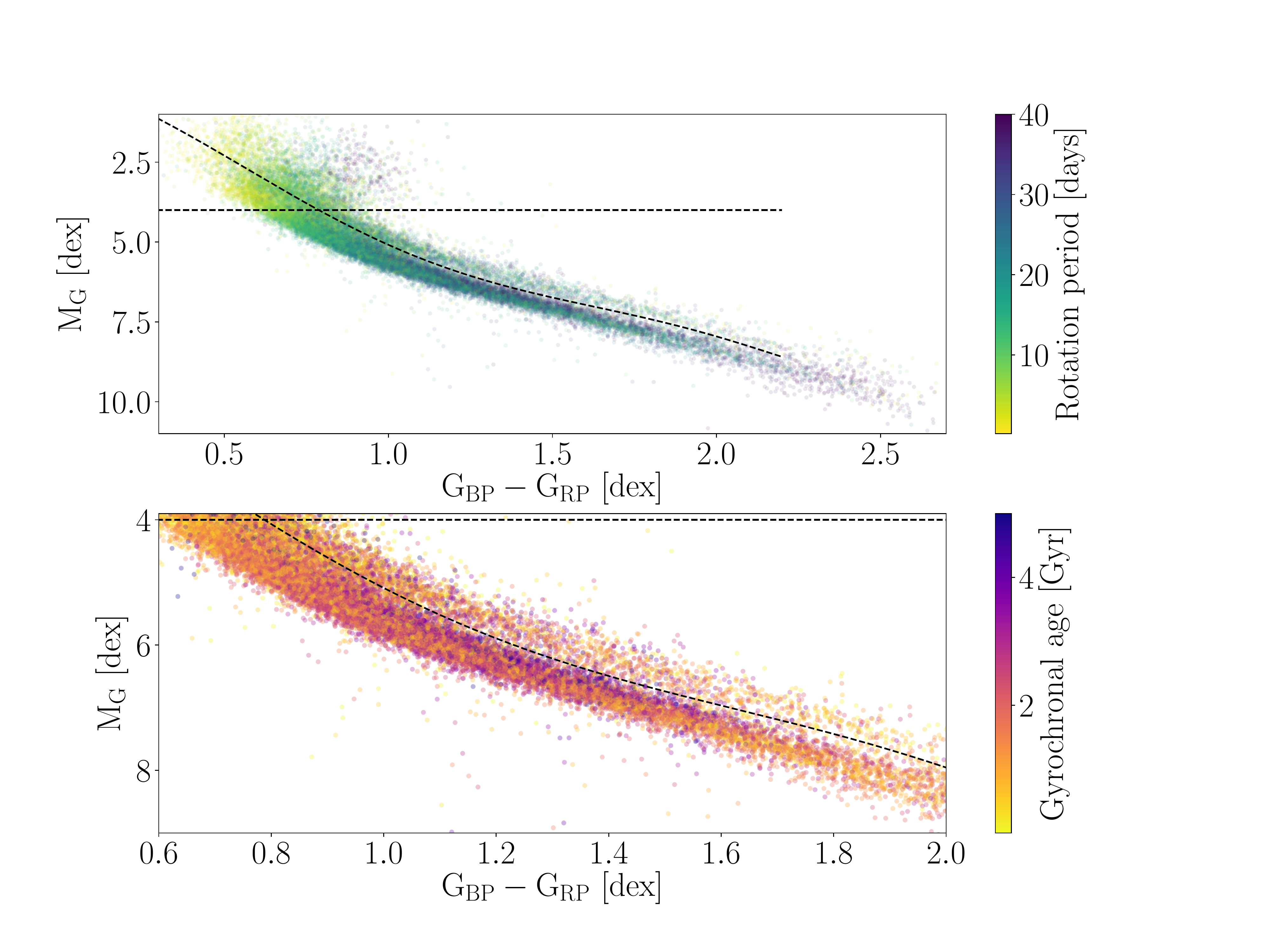}
\label{fig:age_gradient}
\end{figure}

Photometric binaries and subgiants were removed from the \mct\ sample by
applying cuts to the color-magnitude diagram (CMD), shown in figure
\ref{fig:age_gradient}.
A 6th-order polynomial was fit to the main sequence and raised by 0.27 dex to
approximate the division between single stars and photometric binaries (shown
as the curved dashed line in figure \ref{fig:age_gradient}).
All stars above this line were removed from the sample.
Potential subgiants were also removed by eliminating stars brighter than 4th
absolute magnitude in \gaia\ G-band.
This cut also removed a number of main sequence F stars from our sample,
however these hot stars are not the focus of our gyrochronology study since
their small convective zones inhibit the generation of a strong magnetic
field.
The removal of photometric binaries and evolved/hot stars reduced the total
sample of around 34,000 stars by almost 10,000.

The rotation periods of the dwarf stars in the \mct\ sample are shown on a
\gaia\ color-magnitude diagram (CMD) in the top panel of figure
\ref{fig:age_gradient}.
In the bottom panel, the stars are colored by their gyrochronal age,
calculated using the \citet{angus2019} gyrochronology relation.
The stars with old gyrochronal ages, plotted in purple hues, predominantly lie
along the upper edge of the MS, where stellar evolution models predict old
stars to be, however the majority of these `old' stars are bluer than \gcolor\
$\sim$ 1.5 dex.
The lack of gyrochronologically old M dwarfs suggests that either old M dwarfs
are missing from the \mct\ catalog, or the \citet{angus2019} gyrochronology
relation under-predicts the ages of low-mass stars.
Given that lower-mass stars stay active for longer than higher-mass stars
\citep[\eg][]{west2008, newton2017, kiman2019}, and are therefore more likely
to have measurable rotation periods at old ages, the latter scenario seems
likely.
However, it is also possible that the rotation periods of the oldest early M
dwarfs are so long that they are not measurable with Kepler data.
\racomment{Ground-based rotation period measurements of early M dwarfs
(spectral types earlier than $\sim$ M2.5, \teff\ $\gtrsim$ 3500 K) indicate
there is a $\sim$ 80 day upper limit to their rotation periods
\citep{newton2016, newton2018}, which is longer than the longest rotation
periods measured for the early M dwarfs in the \mct\ sample (around 50 days).
}

The apparent lack of old gyro-ages for M dwarfs in figure
\ref{fig:age_gradient} may be caused by a combination of ages being
underestimated by a poorly calibrated model, and rotation period detection
bias.
The \citet{angus2019} gyrochronology relation is a simple polynomial model,
fit to the period-color relation of Praesepe.
Inaccuracies at low masses are a typical feature of empirically calibrated
gyrochronology models since there are no (or at least very few) old M dwarfs
with rotation periods and the models are poorly calibrated for these stars.

The {\tt Pyia} \citep{price-whelan2018} and {\tt astropy} \citep{astropy2013,
astropy2018} {\it Python} packages were used to calculate velocities for the
\mct\ sample.
{\tt Pyia} calculates velocity samples from the full \gaia\ uncertainty
covariance matrix via Monte Carlo sampling, thereby accounting for the
covariances between \gaia\ positions, parallaxes and proper motions.
Stars with negative parallaxes or parallax signal-to-noise ratios less than 10
(around 3,000 stars), stars fainter than 16th magnitude \racomment{in \gaia\
G-band} (200 stars), stars with absolute \vb\ uncertainties greater than 1
\kms\ (1000 stars), and stars with galactic latitudes greater than 15\degrees\
(5500 stars, justification provided in the appendix) were removed from the
sample.
Finally, we removed almost 2000 stars with rotation periods shorter than the
main population of periods, since this area of the period-\teff\ diagram is
sparsely populated.
We removed these rapid rotators by cutting out stars with gyrochronal ages
less than 0.5 Gyr \citep[based on the][gyro-model]{angus2019}, because a 0.5
Gyr gyrochrone\footnote{A gyrochrone is a gyrochronological isochrone, or a
line of constant age in period-\teff, or period-color space.} traces the
bottom edge of the main population of rotation periods.
\racomment{These rapid rotators are probably either very young stars, or
synchronized binaries (see section \ref{sec:gap}).}
After these cuts, around
13,000 stars out of the original $\sim$34,000 were included in the sample.

\section{Results and Discussion}
\label{sec:results}

\subsection{The period-\teff\ relations, revealed}
\label{sec:the_reveal}

\begin{figure}
  \caption{
    Top: Rotation period vs effective temperature for stars in the \mct\
    sample, colored by the velocity dispersions of stars calculated over a
    grid in $\log_{10}$(period) and \teff\ (this grid causes the quantized
    appearance).
    Black lines show gyrochrones from a gyrochronology model that projects the
    rotation-color relation of
    Praesepe to longer rotation periods over time \citep{angus2019}.
    These gyrochrones do not appear to reflect the evolution of field stars at
    long rotation periods/old ages because they do not trace lines of constant
    velocity dispersion.
    Gyrochrones are plotted at 0.5, 1, 1.5, 2, 2.5, 4 and 4.57 Gyr (Solar age)
    in both top and bottom panels.
    Bottom: Same as top panel with rotation period vs {\it mass}
    \citep[from][]{berger2020}.
    White lines show gyrochrones from a model that includes mass and
    age-dependent angular momentum transport between the core and envelope
    \citep{spada2019}.
    Qualitatively, these gyrochrones reflect the evolution of field
    stars at long rotation periods/old ages: they trace lines of constant
    velocity dispersion by reproducing periods of `stalled' surface rotational
    evolution for K-dwarfs.
    The data used to create this figure is available in table \ref{tab:data}.
}
  \centering
    \includegraphics[width=1\textwidth]{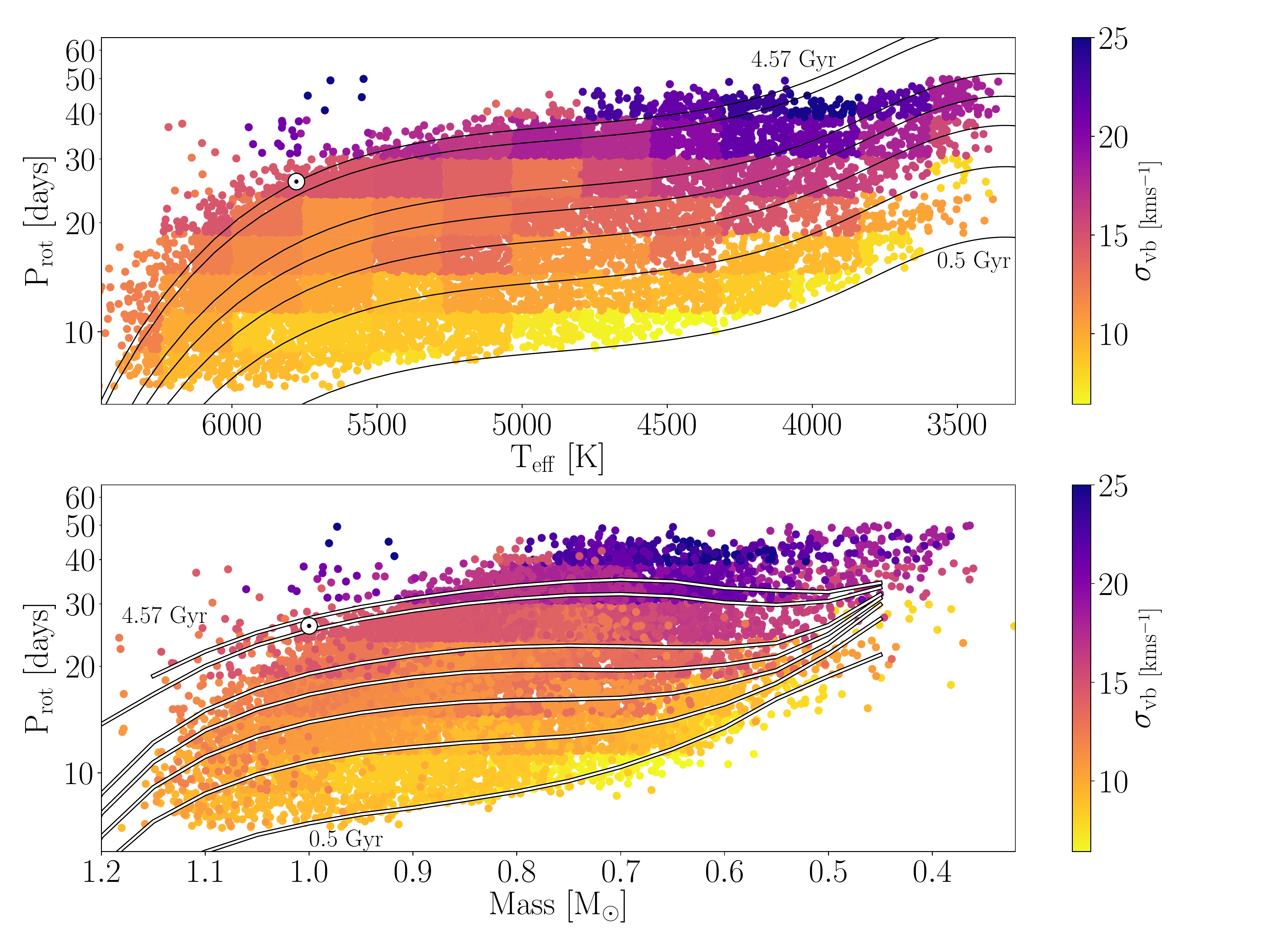}
\label{fig:vplot}
\end{figure}

To explore the relationship between rotation period, effective temperature
(\teff ) and velocity dispersion, we calculated \sigmavb \footnote{\sigmavb\
was calculated as 1.5$\times$ the median absolute deviation of velocities, to
mitigate sensitivity to outliers.} for populations of stars with similar
rotation periods and temperatures, and presumed similar age.
The top panel of figure \ref{fig:vplot} shows rotation period versus effective
temperature for the \mct\ sample, coloured by \sigmavb, where \sigmavb\ was
calculated for groups of stars over a grid in $\log_{10}$(period) and
temperature.
If we assume that mass dependent heating does not strongly affect this sample
and \vb\ at low galactic latitudes is an unbiased tracer of \vz, then \vb\
velocity dispersion can be interpreted as an age proxy, and stars plotted in a
similar color in figure \ref{fig:vplot} are similar ages.
In the appendix of this paper, we show that this assumption appears valid for
stars with Galactic latitude $<$ 15\degrees.

Overall, figure \ref{fig:vplot} shows that velocity dispersion increases with
rotation period across all temperatures, implying that rotation period
increases with age, as expected.
This result is insensitive to the choice of bin position and size.
Black lines show gyrochrones from the \citet{angus2019} gyrochronology model,
which projects the rotation-color relation of Praesepe to longer rotation
periods over time.
These gyrochrones are plotted at 0.5, 1, 1.5, 2, 2.5, 4 and 4.57 (Solar age)
Gyr.
At the youngest ages, these gyrochrones describe the data well: the palest
yellow (youngest) stars with the lowest velocity dispersions all fall close to
the 0.5 Gyr gyrochrone.
However, although the 0.5 Gyr and 1 Gyr gyrochrones also trace constant
velocity dispersion/age among the field stars, by 1.5 Gyr the gyrochrones
start to {\it cross} different velocity dispersion regimes.
For example, the 1.5 Gyr gyrochrone lies on top of stars with velocity
dispersions of around 10-11 kms$^{-1}$ at 5000-5500K and stars with $\sim$15
\kms\ velocity dispersions at 4000-4500K.
The gyrochrones older than 1.5 Gyr also cross a range of velocity dispersions.
If these were true isochrones they would follow lines of constant velocity
dispersion.
At ages older than around 1 Gyr, it appears that gyrochrones should have a
more flattened, or even inverted, shape in rotation period-\teff\ space than
these Praesepe-based models.

The bottom panel of figure \ref{fig:vplot} shows velocity dispersion as a
function of rotation period and {\it mass}, \citep[from][]{berger2020}, with
gyrochrones from the \citet{spada2019} model shown in white.
These gyrochrones are \racomment{plotted for the same ages described above.}
Each point plotted in the top panel also appears in the bottom panel with the
same color.
Because velocity dispersion was calculated in bins of \teff, not mass, bin
outlines are clearly visible in the top panel but appear smeared-out in the
bottom panel.
In the bottom panel of figure \ref{fig:vplot}, the \citet{spada2019} models
{\it do} trace lines of constant velocity dispersion, and reproduce the trends
in the data at all ages.
These models qualitatively agree with the data and reproduce the apparent
flattening and inversion in the rotation period-\teff/mass relations.

The results shown in figure \ref{fig:vplot} indicate that stars of spectral
type ranging from late G to late K ($\sim$5500-3500 K) follow a braking law
that changes over time.
In particular, the relationship between rotation period and effective
temperature appears to flatten out and eventually invert.
These results provide further evidence for `stalled' surface rotational
evolution of K dwarfs, like that observed in open clusters \citep{curtis2019}
and reproduced by models that vary angular momentum transport between stellar
core and envelope with time and mass \citep{spada2019}.

The velocity dispersions of stars in the \mct\ sample, shown in figure
\ref{fig:vplot}, provide the following picture of rotational evolution.
\racomment{At young ages \citep[younger than around 1 Gyr but still old enough
to be on the main sequence and have transitioned from the `C' sequence to the
`I' sequence ][]{barnes2003}, stellar rotation period {\it decreases} with
{\it increasing} mass.}
This is likely because lower-mass stars with deeper
convection zones have stronger magnetic fields, larger Alfv\'en radii and
therefore experience greater angular momentum loss rate
\citep[\eg][]{schatzman1962, kraft1967, parker1970, kawaler1988,
charbonneau2010, matt2012, matt2015}.
According to the \citet{spada2019} model, the radiative cores and convective
envelopes of stars are decoupled at these young ages, \ie\ transportation of
angular momentum from the surface to the core of the star is reduced, so the
surface slows down due to wind-braking but the core keeps spinning rapidly.
According to the data presented in figure \ref{fig:vplot}, at intermediate
ages, the rotation periods of K dwarfs appear {\it constant} with mass, and at
late ages rotation period {\it increases} with {\it increasing} mass.
The interpretation of this, according to the \citet{spada2019} model, is that
lower-mass stars are still braking more efficiently at these intermediate and
old ages \racomment{compared to higher-mass stars,} but their cores are more
tightly coupled to their envelopes, allowing angular momentum transport
between the two layers.
Angular momentum surfaces \racomment{from the core} and prevents the stellar
envelopes from spinning-down rapidly, and this effect is strongest for late
K-dwarfs with effective temperatures of $\sim$4000-4500K and masses
$\sim$0.5-0.7 M$_\odot$.

A period of core-envelope decoupling in the evolution of cool dwarfs has been
explored in theoretical models for decades \citep[\eg][]{endal1981,
macgregor1991, denissenkov2010, gallet2013}.
In such models, the angular momenta of the radiative core and convective
envelope are permitted to evolve separately once a radiative core develops on
the pre-main sequence.
A decoupled core and envelope is required to reproduce observations of young
clusters and star forming regions \citep[\eg][]{irwin2007, bouvier2008,
denissenkov2010, spada2011, reiners2012} and has become an established element
of theoretical gyrochronology models.
During this phase, angular momentum transport between the radiative core and
the convective envelope is reduced.
Over time, models increase the efficiency of angular momentum transport
between the core and envelope in order to reproduce the close-to solid body
rotation observed for the Sun \citep[\eg][]{thompson1996}.
The core-envelope coupling timescale affects the predicted surface rotation
periods of young and intermediate-age stars and is usually constrained using
observations of open clusters.
The \citet{lanzafame2015} gyrochronology model uses a mass-dependent
core-envelope coupling timescale, and \citet{spada2019} fit this model to
open-cluster observations, including new rotation period measurements for
K dwarfs in the NGC 6811 cluster \citep{curtis2019}.
A similar mass-dependent core-envelope coupling timescale was also found to
explain the observed lithium depletion trends in open clusters by an
independent study \citep{somers2016}.
Although variable angular momentum transport between the surfaces and cores of
stars has been an essential ingredient of stellar evolution models for
decades, the transport mechanism is still unknown.
Among the proposed mechanisms are magneto-hydrodynamical waves resulting from
various magnetic field geometries, and gravity waves \citep[see,
\eg][]{charbonneau1993, ruediger1996, spruit2002, talon2003, spada2010,
brun2011, oglethorpe2013}.


\racomment{In the top panel of figure \ref{fig:vplot}, the \citet{angus2019}
gyrochronology model is plotted for comparison with the data.
This model was chosen as an example but there are many other empirical
gyrochronology models we could have used.
All the available empirical gyrochronology models are similar in essence to
the \citet{angus2019} relation and, like that relation, none is able to
reproduce the observed velocity dispersions, or capture the evolving shape of
the period-\teff\ relations.

The \citet{angus2019} relation is a new fully-empirical gyrochronology
relation, calibrated using recently measured rotation periods for members of
the Praesepe cluster, which extend down to early M dwarfs \citep{rebull2017,
douglas2017}.
These are the oldest cluster M dwarfs with measured photometric rotation
periods and the \citet{angus2019} model therefore encapsulates the behavior of
these cool stars at Praesepe age.
Most semi-empirical spin-down models predict that, once the rotation periods of
stars converge to a tight sequence, they approximately spin down with a common
braking index \citep[e.g., Fig. 5 in][]{gallet2015}.
So, if stellar rotation periods have fully converged by the age of Praesepe,
as the observations suggest they have \citep{douglas2019}, it is appropriate
to project the Praesepe sequence forward in time with a single braking index
that is common to all stars and constant in time.
However, observations of low-mass stars in older clusters \citep[\eg, NGC
6811, NGC 752, and Ruprecht 147][Curtis \etal\ 2020, submitted]{curtis2019,
agueros2018} now demonstrate that the assumptions adopted in the
semi-empirical models which support a common braking index must be invalid, or
inaccurate in some quantitative way (for example, the core-envelope coupling
timescale is inaccurately calibrated).

There are many other empirical gyrochronology models we could have chosen to
compare with our data.
For example, the \citet{barnes2003, barnes2007, mamajek2008, meibom2011,
angus2015} relations all have a functional form that was first introduced by
\citet{barnes2003}: $\mathrm{P_{rot}} = t^n~a(B-V - c)^b$, where n, a, b and c
are free parameters, and B and V are photometric magnitudes.
These relations have been widely used in the literature, and are similar to
the \citet{angus2019} relation in that they consist of a color-dependent term
multiplied by an age-dependent term, \ie\ they are separable in color and age.
None of these relations follow lines of constant velocity dispersion in figure
\ref{fig:vplot} because they have a universal power-law index, $n$
\citep[$\approx$ 0.5,][]{skumanich1972}, which applies to stars of all masses
and ages.
They do not have the flexibility to capture the evolving period-color
relationship seen in the data.
The \citet{barnes2010} model has a different functional form: it uses Rossby
number ($Ro = P_\mathrm{rot}/\tau$, where $\tau$ is the convective turnover
time) to encode the mass-dependent evolution of rotation periods:
$\frac{dP_\mathrm{rot}}{dt} \propto 1/Ro$.
This simple relation neatly reproduces the rotation periods of stars in the
slow-rotation sequences of young ($\sim$100-700 Gyr) clusters, however, like
the other empirical gyrochronology models, it does not reproduce the observed
trends seen in the velocity dispersion data here.
The \citet{barnes2010} model has a similar period-\teff\ relation to the
Praesepe-based \citet{angus2019} model, because the period-\teff\ relation of
Praesepe roughly follows a line of constant Rossby number.
Consequently, like the \citet{angus2019} model, the \citet{barnes2010} model
in its current form does not have the flexibility to capture the mass and
time-variable rotational evolution revealed by the data explored here.


It is not surprising that the \citet{angus2019} model, or other gyrochronology
models do not reproduce the data, since tensions with Skumanich-like spin-down
have been revealed whenever new rotation periods of benchmark stars have
become available \citep{angus2015, vansaders2016, metcalfe2019, agueros2018,
curtis2019}.
With the influx of new stellar rotation periods provided by \kepler/\ktwo,
\tess, and ground-based facilities, our understanding of spin-down is changing
rapidly.
New data are revealing flaws in not just the calibration of but also the
functional forms of old models.
We have shown that velocity dispersion data provides support for more
sophisticated models that incorporate additional physical effects \citep[\eg,
core-envelope coupling][]{spada2019} over the simple empirical models used
today.
Alternatively, more flexible empirical frameworks, like a Gaussian process
model, could capture everything we now know about stellar spin-down.
}

\racomment{

\subsection{The mass-dependence of magnetic activity lifetimes}
\label{sec:activity}

The mass dependence of magnetic activity lifetimes has been demonstrated
previously \racomment{for M dwarfs} \citep[\eg][]{west2008, reiners2008,
west2009, newton2017, kiman2019} and, if the detectability of a rotation
period is considered to be a magnetic activity proxy, then our results provide
further evidence for a mass-dependent activity lifetime.

In order to measure a rotation period at all, there must be some magnetically
active regions (either bright plages or dark spots), of a reasonable size on
the surface of a star.
In other words, some minimum magnetic activity level is presumably required to
produce coherent light curve-variability, from which a rotation period can be
confidently measured.
The relatively sharp upper-edge of the rotation period distribution seen in
figure \ref{fig:vplot} may therefore be caused by a finite active lifetime of
stars, \ie\ stars older than a critical age are no longer active enough to
produce periodic, high-amplitude variability.
In figure \ref{fig:vplot}, the velocity dispersions of stars along the upper
edge of the rotation period distribution increase with decreasing mass,
indicating that magnetic activity lifetimes are mass-dependent.
Figure \ref{fig:vplot} shows that the populations of stars with the largest
velocity dispersions are cooler than 4500 K.
This implies that most of the oldest stars with detectable rotation periods
are cooler than 4500 K, \ie\ these-low mass stars stay active longer than more
massive stars.
}

\begin{figure}
  \caption{
    Velocity dispersions for the entire \kepler\ field divided by the velocity
    dispersions of stars with measured rotation periods in \mct,
    as a function of effective temperature.
    A larger ratio indicates that the overall \kepler\ field is older, on
    average, than stars in the \mct\ catalog.
    As this ratio approaches unity the two populations have similar kinematic
    ages.
    The large ratio for the hottest stars indicates that G dwarfs become
    inactive at young ages.
    This ratio approaches unity at low temperatures, showing that K and early
    M dwarf rotation periods are measurable over a large range of ages.
}
  \centering
    \includegraphics[width=1\textwidth]{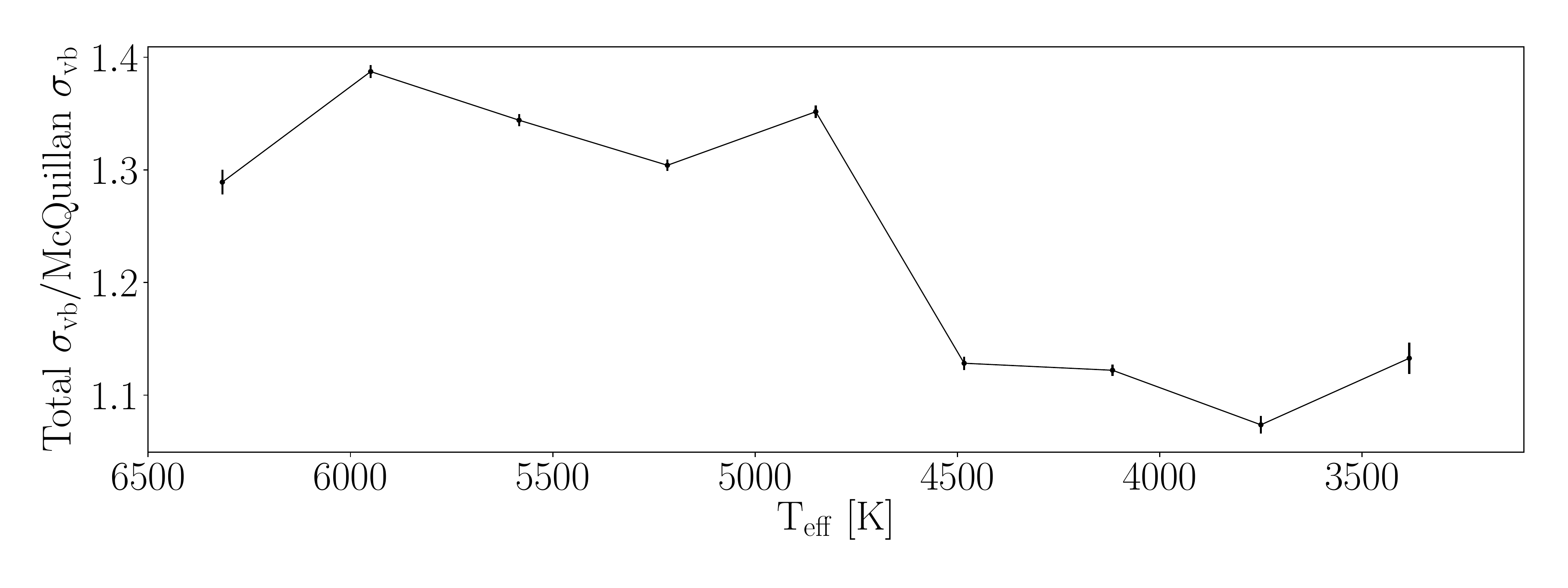}
\label{fig:compare}
\end{figure}

\racomment{
    To investigate this idea further, we compared the velocity dispersions of
stars with {\it measured rotation periods} to the velocity dispersions of the
entire \kepler\ sample.
If stars with measured rotation periods are more magnetically active, and
younger, than stars without rotation period measurements, they should also
have smaller velocity dispersions.
To test this theory, we calculated the velocity dispersions for all stars in
the \kepler\ field (after removing visual binaries, subgiants, stars fainter
than 16th magnitude, and high Galactic latitude stars, following the method
described in section \ref{sec:method}).
We then compared these velocity dispersions, as a function of
\teff, to the velocity dispersions of stars with {\it measured} rotation
periods \citep[\ie\ stars that appear in table 1 of][]{mcquillan2014}.
We show the ratio of total \sigmavb\ to McQuillan-\sigmavb\ as a
function of \teff\ in figure \ref{fig:compare}.}
A larger ratio means the rotating star sample is {\it younger}, on average,
than the overall \Kepler\ population, and a ratio of 1 means that the rotating
stars have the {\it same} age distribution as the overall Kepler sample.
Figure \ref{fig:compare} shows that this ratio is largest for G stars and
approaches unity for K and early M dwarfs.
This indicates that the G stars with detectable rotation periods are, on
average, {\it younger} than the total population of G stars in the Kepler
field.
On the other hand, the late K and early M dwarfs with detectable rotation
periods have a similar age distribution to the overall Kepler population which
suggests that \racomment{most of}
the oldest K and M dwarfs are represented in
the \mct\ sample.
This result bolsters the evidence that M dwarf rotation periods are measurable
at older ages than G dwarf rotation periods and that G stars become
magnetically inactive and have fewer active surface regions {\it at a younger
age than M dwarfs}.

\subsection{Synchronized binaries and the \kepler\ period gap}
\label{sec:gap}

\begin{figure}
  \caption{
      Top: rotation period vs. effective temperature for stars in the \mct\
    sample, separated into three groups. Blue circles
      show stars with rotation periods longer than the
    period gap, orange squares show stars with rotation periods shorter than
    the gap, but longer than the lower edge of the main rotation period
    distribution, and green triangles show stars with rotation periods shorter
    than this lower edge.
    Stars were separated into these three groups using \citet{angus2019}
    gyrochronology models, with the scheme shown in the legend.
    Bottom: the velocities of these groups of stars (in the direction of
    Galactic latitude, $b$) are shown as a function of rotation period.
    Only stars cooler than 5000 K are plotted in the bottom panel in order to
    isolate populations above and below the period gap, which only extends up
    to temperatures of $\sim$4600 K.
    The black line indicates the velocity standard deviation as a function of
    period.
}
  \centering
    \includegraphics[width=1\textwidth]{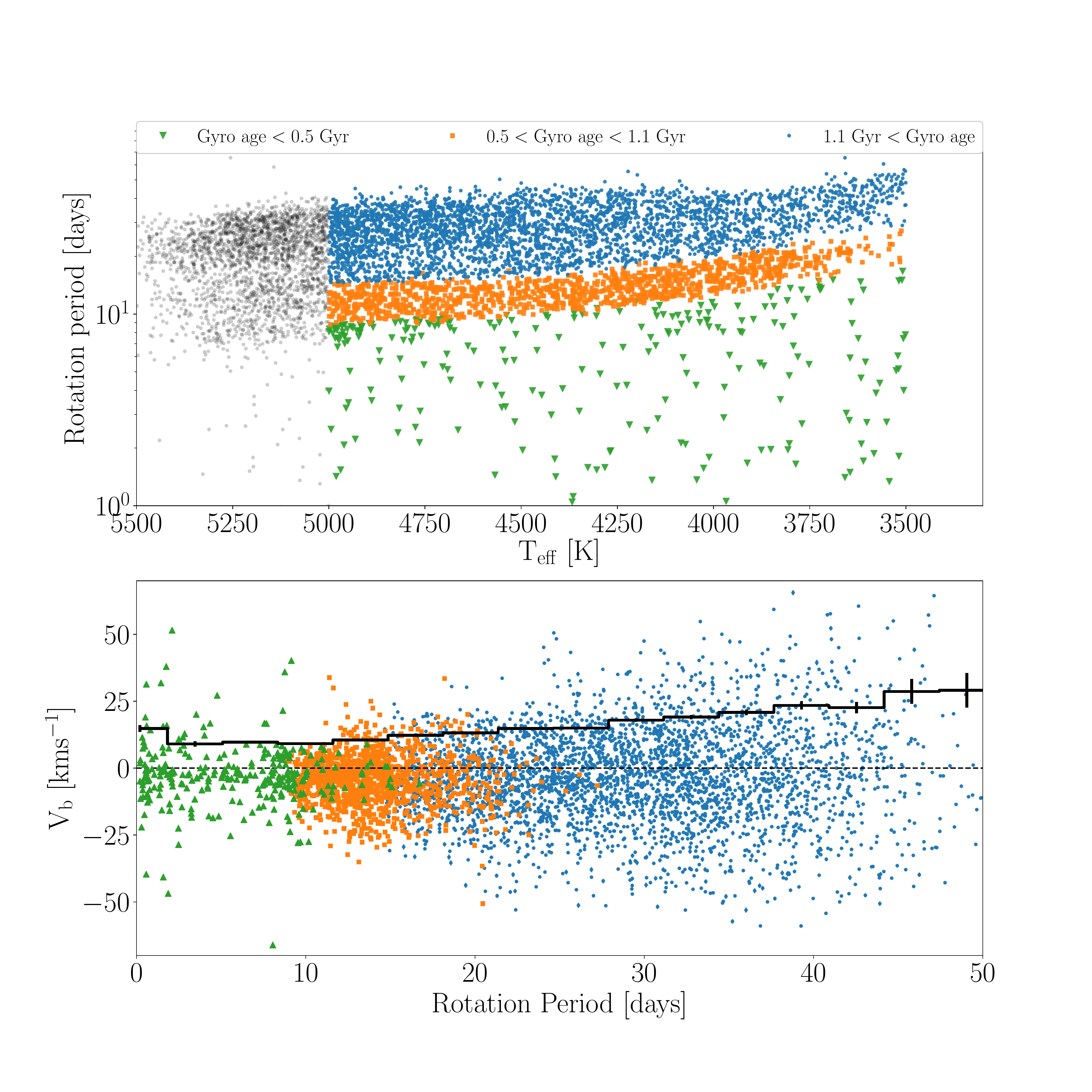}
\label{fig:gap}
\end{figure}

In this section, we explored the kinematic properties of the \mct\ sample in
more detail, investigating the velocity dispersions of stars on either side of
the \kepler\ period gap, and identifying rapidly rotating stars that may be
synchronized binaries.

There is a sharp gap in the population of rotation periods (often called the
\kepler\ period gap), which lies just above the 1 Gyr gyrochrone in the upper
panel of figure \ref{fig:vplot}, whose origin is unknown and is the subject of
much speculation \citep{mcquillan2014, davenport2017, davenport2018,
reinhold2019, reinhold2020}.
This gap was first identified by \mct, and roughly follows a line of constant
gyrochronal age of around 1.1 Gyr \citep[according to the][gyrochronology
relation]{angus2019}.
Several explanations for the gap's origin have been proposed, including a
discontinuous star formation history \citep{mcquillan2013, davenport2017,
davenport2018} and a change in magnetic field structure causing a brief period
where rotational variability is reduced and rotation periods cannot be
measured \citep{reinhold2019, reinhold2020}.

The top panel of figure \ref{fig:vplot} suggests that the \citet{angus2019},
Praesepe-based gyrochronology model is valid below the gap but not above.
Gyrochrones follow lines of constant velocity dispersion below the gap, but
{\it cross} lines of constant velocity dispersion above the gap.
This phenomenon is robust to the choice of bin size and position.
Although we do not provide an in-depth analysis here (and more data may be
needed to confirm a connection) these data suggest that the gap may indeed
separate a young regime where stellar cores are decoupled from their envelopes
from an old regime where these layers are more tightly coupled.
If so, this could indicate that
\racomment{these phenomena are related, \ie\
the process that is responsible for changing the shape of gyrochrones in
rotation-\teff\ space is related to the process that produces the gap.}

An alternate explanation for the gap is that the \mct\ sample contains two
distinct stellar populations: one young and one old.
If so, the kinematic properties of stars above and below the gap are likely to
be distinctly different.
The bottom panel of figure \ref{fig:gap} shows the velocity dispersions of
stars in the \mct\ sample, with stars subdivided into three groups: those that
rotate more quickly than the main rotation period population (green
triangles), those with rotation periods shorter than the gap (orange squares),
and those with rotation periods longer than the gap (blue circles).
Stars were separated into these three groups using \citet{angus2019}
gyrochronology model, according to the scheme shown in the legend.
Only stars cooler than 5000 K are included in the bottom panel in order to
isolate populations above and below the period gap, which only extends up to a
temperature of $\sim$4600 K in our sample, although \citet{davenport2017}
found that the gap extends to temperatures as hot as 6000 K.
In general, velocity dispersion increases with rotation period because both
quantities increase with age.
Previously, only the overall velocity dispersions of all stars above and below
the gap have been compared, leading to the assumption that these groups belong
to two distinct populations \citep{mcquillan2014}.
However, figure \ref{fig:gap} shows a smooth increase in velocity dispersion
with rotation period across the gap (from orange squares to blue circles),
suggesting that these groups are part of the same Galactic population.
This observation does not rule out the possibility that a brief cessation of
star formation in the Solar neighborhood, around one Gyr ago, may have caused
this gap, however.

In the final part of our analysis, we investigated the potential for using
kinematics to identify synchronized binaries in the \mct\ sample.
Synchronized binaries are pairs of stars whose rotation periods are equal to
their orbital period.
Since synchronization appears to happen at rotation periods of 7 days or
shorter \citep{simonian2019}, and most isolated stars have rotation periods
longer than 7 days, the rotation periods of synchronized binaries are likely
to be {\it shorter} than they would be if they were isolated stars.
For this reason, their rotation periods do not reflect their ages and the
gyrochronal age of a synchronized binary is likely to be much younger than the
true age of the system.
Synchronized binaries are therefore a source of contamination for
gyrochronology and should be removed from samples before performing a
gyrochronal age analysis.
Figure \ref{fig:gap} shows that some of the most rapidly rotating stars in the
\mct\ sample have relatively large absolute velocities, indicating that they
are likely synchronized binaries.
For this reason, the velocity dispersions of stars with rotation periods
shorter than the lower edge of the rotation period distribution (green
triangles in figure \ref{fig:gap}) are not significantly smaller than the,
presumed older, orange-colored stars.
In general, stars with rotation periods less than $\sim$10 days have an
increased chance of being synchronized binaries.
This result is in agreement with a recent study which found that a large
fraction of photometric binaries were rapid rotators, and the probability of a
star being a synchronized binary system substantially increased below rotation
periods of around 7 days \citep{simonian2019}.
We caution users of rotation period catalogs that rapid rotators with large
absolute velocities should be flagged as potential synchronized binaries
before applying any gyrochronal analysis.

\section{Conclusions}

In this paper, we used the \vb\ velocity dispersions of stars in the \mct\
catalog to explore the evolution of stellar rotation period as a function of
effective temperature and age.
Our conclusions are as follows:
\begin{itemize}
\item{{\bf Spin-down rate doesn't always increase with decreasing mass for K
    dwarfs.}
Although at young ages, rotation period is anti-correlated with \teff\ (as
seen in many young open clusters, including Praesepe), at intermediate ages the
relation flattens out and K dwarfs of different masses rotate at the same
rate.
At old ages, cooler K dwarfs spin more rapidly than hotter K dwarfs of the
same age.
        \racomment{ While most empirical gyrochronology models calibrated to
        date can broadly reproduce the rotation periods of young
        ($\sim$100-700 Gyr) cluster stars, they do not match the rotational
        behavior of intermediate-age and old ($\gtrsim$1 Gyr) K dwarfs.
In addition, their simple functional forms, with a universal age-rotation
        power-law index, are not flexible enough to capture rotational
        evolution at all ages and masses.
This is because they have been calibrated with young open clusters: stellar
        spin down is relatively straightforward at young ages, and can often
        be represented with a color and age-separable power-law relation.
Until recently, more complex rotational evolution models were not needed.
We now know however, from new observations of intermediate-age clusters, and
        the kinematic data presented here, that these empirical models are not
        flexible enough to reproduce the mass and time-dependent spin-down
        rate of cool dwarfs at old ages.
We advocate for the adoption of more flexible models, like semi-parametric
        Gaussian process models, in the future.
        }
        }

\item{{\bf Variable core-envelope coupling may be the cause.} We showed that
the period--\teff\ relations change shape over time in a way that
qualitatively agrees with theoretical models which include mass and
time-dependent core-envelope angular momentum transport \citep{spada2019}.}

\item{{\bf Low-mass stars stay active longer.}
We found that the oldest stars in the \mct\ catalog are cooler than 4500
K, which agrees with previous results which show that lower-mass stars remain
        active for longer, allowing their rotation periods to be measured at
        older ages.}

\item{{\bf The \kepler\ period gap may be related to core-envelope coupling.}
We speculated that the rotation period gap \citep{mcquillan2014} may separate
a young regime where stellar rotation periods decrease with increasing mass
from an old regime where periods increase with increasing mass, however more
data are needed to provide a conclusive result.
The velocity dispersions of stars increase smoothly across the rotation period
gap, indicating that the gap does not separate two distinct stellar
populations.}

\item{{\bf Rapidly rotating stars with large absolute velocities may be
synchronized binaries.}
We used kinematics to indicate that there is a population of
synchronized binaries with rotation periods less than around 10 days.}


\end{itemize}

We thank the anonymous referee for their comments which greatly improved this
manuscript.
We also would like to thank Suzanne Aigrain for providing thoughtful insight
that improved the paper.
This work was partly developed at the 2019 KITP conference `Better stars,
better planets'.
This research was supported in part by the National Science Foundation under
Grant No. NSF PHY-1748958.
Parts of this project are based on ideas explored at the Gaia sprints at the
Flatiron Institute in New York City, 2016 and MPIA, Heidelberg, 2017.
This work made use of the gaia-kepler.fun crossmatch database created by Megan
Bedell.
T.A.B. acknowledges support by a NASA FINESST award (80NSSC19K1424).

Some of the data presented in this paper were obtained from the Mikulski
Archive for Space Telescopes (MAST).
STScI is operated by the Association of Universities for Research in
Astronomy, Inc., under NASA contract NAS5-26555.
Support for MAST for non-HST data is provided by the NASA Office of Space
Science via grant NNX09AF08G and by other grants and contracts.
This paper includes data collected by the Kepler mission. Funding for the
\Kepler\ mission is provided by the NASA Science Mission directorate.

This work has made use of data from the European Space Agency (ESA) mission
{\it Gaia} (\url{https://www.cosmos.esa.int/gaia}), processed by the {\it
Gaia} Data Processing and Analysis Consortium (DPAC,
\url{https://www.cosmos.esa.int/web/gaia/dpac/consortium}).
Funding for the DPAC has been provided by national institutions, in particular
the institutions participating in the {\it Gaia} Multilateral Agreement.

\section{Appendix A: Validating \vb\ dispersion as an age proxy}
\label{sec:mass-dependent-heating}

The conclusions drawn in this paper depend on the assumption that velocity
dispersion in the direction of Galactic latitude (\sigmavb) can be used as an
age proxy.
There are two main reasons however, why \vb\ velocity dispersion may {\it not}
be a good age proxy.
Firstly, mass-dependent heating may act on the sample, meaning that velocity
dispersion depends on both age and mass.
Secondly, since stars in the \kepler\ field have a range of Galactic
latitudes, using \vb\ as a stand-in for \vz\ may not be equally valid for all
stars, and introduce a velocity bias for high latitude stars (which are more
likely to be cooler and older).
In this section we demonstrate that neither of these problems seem to be a
significant issue for our data.

In order to establish whether \sigmavb\ can be used as an age proxy, we
searched for signs of mass-dependent heating within the \kepler\ field.
Mass-dependent dynamical heating may result from lower-mass stars experiencing
greater velocity changes when gravitationally perturbed than more massive
stars.
It has not been unambiguously observed in the galactic disk because of the
strong anti-correlation between stellar mass and stellar age.
Less massive stars do indeed have larger velocity dispersions, however they
are also older on average.
This mass-age degeneracy is highly reduced in M dwarfs because their
main-sequence lifetimes are longer than the age of the Universe, and no
evidence for mass-dependent heating has previously been found in M dwarfs
\citep[\eg][]{faherty2009, newton2016}.

To investigate whether mass-dependent heating could be acting on the \kepler\
sample, we selected late K and early M dwarfs observed by both \kepler\ and
\gaia, whose MS lifetimes exceed around 11 Gyr and are therefore
representative of the initial mass function.
We could not perform this analysis on the \mct\ sample, because it only
includes stars with {\it detectable} rotation periods, and since lower-mass
stars stay active for longer it is likely that it contains a strong mass-age
correlation.
We selected all \kepler\ targets with dereddened \gaia\ \gcolor\ colors
greater than 1.2 (corresponding to an effective temperature $\lesssim$
4800 K) and absolute \gaia\ $G$-band magnitudes $>$ 4.
We also eliminated photometric binaries by removing stars above a 6th order
polynomial, fit to the MS on the \gaia\ CMD (similar to the one shown in
figure \ref{fig:age_gradient}).
We then applied the quality cuts described above in section
\ref{sec:the_data}.
To search for evidence of mass-dependent heating we calculated the (\vb)
velocity dispersion of stars in effective temperature bins.
Sigma clipping was performed at 3$\sigma$ to remove high and low velocity
outliers before calculating the standard deviation of stars in each bin.
These extreme velocity outliers may be very old late K and early M dwarfs, or
they result from using \vb\ instead of \vz, which introduces additional
velocity scatter.

\begin{figure}
  \caption{
      Top: Stellar velocity (\vb) as a function of \teff\ for
      \kepler\ K and M dwarfs.
Vertical lines indicate different \teff-groupings used to calculate velocity
    dispersion.
Pink stars were not included in velocity dispersion calculations as they were
    either removed as outliers during a sigma clipping process, or they lie at
    the sparcely populated, extremely cool end of the temperature range.
    Velocity dispersion and \teff\ are slightly positively correlated, likely
    due to a brightness-related selection bias, indicating that mass-dependent
    heating does not significantly affect low-mass stars in the \kepler\
    field.
}
  \centering
    \includegraphics[width=1\textwidth]{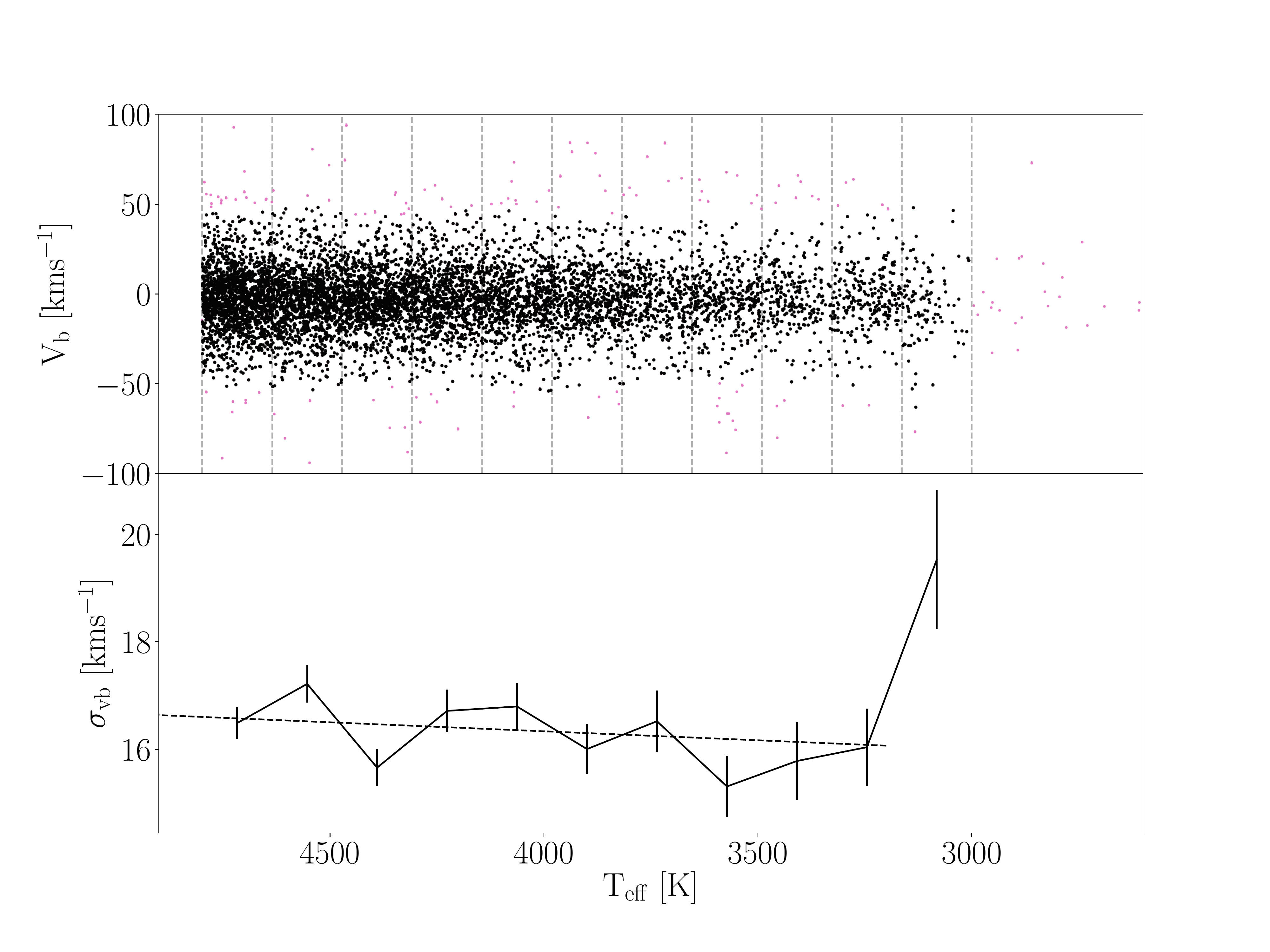}
\label{fig:vb_vs_teff}
\end{figure}

Figure \ref{fig:vb_vs_teff} shows velocity and velocity dispersion as a
function of effective temperature for the K and M \kepler\ dwarf sample.
Velocity dispersion very slightly {\it decreases} with decreasing temperature,
the opposite of the trend expected for mass-dependent heating, however the
slope is only inconsistent with zero at 1.3 $\sigma$.
The sharp uptick in velocity dispersion in the coolest bin is probably noise
caused by the small number of stars in that bin.
This trend may be due to a selection bias: cooler stars are fainter and
therefore typically closer, with smaller heights above the galactic plane and
smaller velocities.
The essential point however, is that we do not see evidence for mass-dependent
heating acting on stars in the \kepler\ field, indicating that velocity
dispersion {\it can} be used as an age proxy (with the caveat that there is
still a chance, albeit a small one, that the opposing effects of the selection
function and mass-dependent heating are working to cancel each other out).
This analysis was performed using \vb\ but we also examined the {\it vertical}
velocities of the 537 stars in this sample with RV measurements.
Again, no evidence was found for mass-dependent heating: the slope of the
velocity dispersion-temperature relation was consistent with zero.

\begin{figure}
  \caption{
This figure demonstrates the variance in the relationship between \vb\ and
    \vz\ for stars in the \kepler\ field, based on the GUMS simulation.
The panels show a kernel density estimator (KDE) (black solid line) for
    the \vz -- \vb\ residuals of stars in the GUMS simulation at four
    different Galactic latitudes.
Blue dashed lines show Gaussian fits to these KDEs.
The distributions are close to Gaussian, with slightly heavy tails.
The standard deviations of the Gaussian fits increase with Galactic latitude.
This figure illustrates how using \vb\ instead of \vz\ artificially
    increases velocity dispersion, especially at high latitudes.
}
  \centering
    \includegraphics[width=1\textwidth]{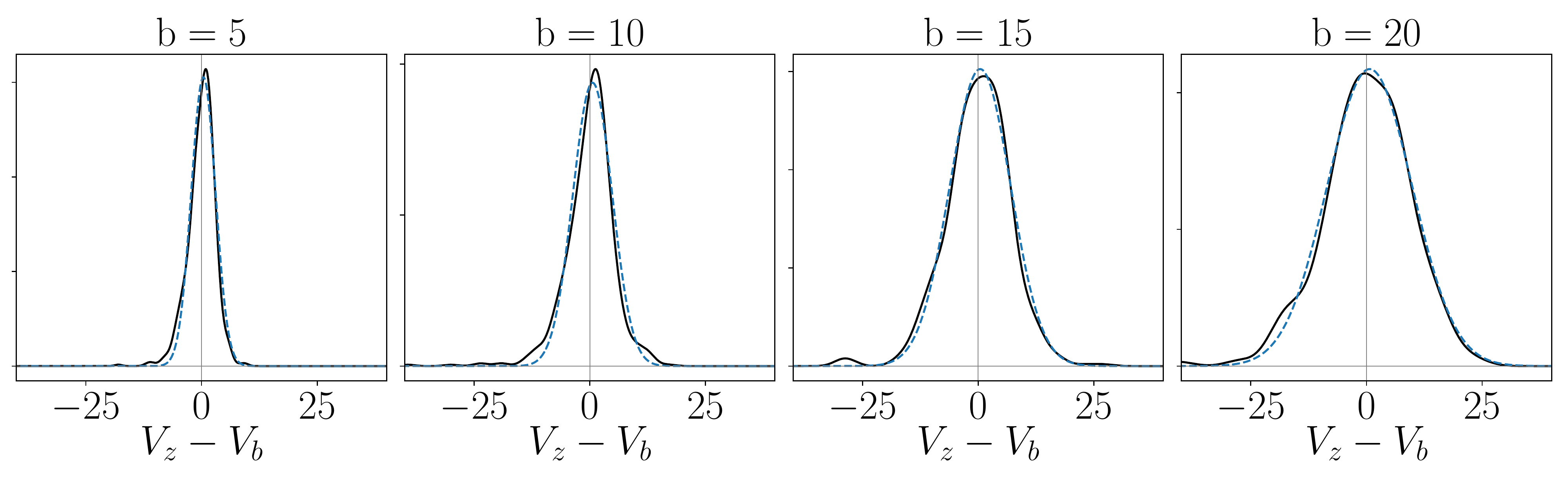}
\label{fig:vb_vz}
\end{figure}

Having found no strong evidence for mass-dependent heating, we next tested
the validity of \vb\ as a proxy for \vz\ in more detail.
At a galactic latitude, $b$, of zero, $v_b=v_z$, however for increasing values
of $b$, this equivalence becomes an approximation that grows noisier with $b$.
To test the validity of the \vb$\sim$\vz\ approximation over a range of
latitudes we downloaded stellar data from the \Gaia\ Universe Model Snapshot
(GUMS) simulation -- a simulated \Gaia\ catalog \citep{robin2012}.
We downloaded stars from four pointings in the \kepler\ field with galactic
latitudes of around 5\degrees, 10\degrees, 15\degrees, and 20\degrees, out to
a limiting magnitude of 16 dex, and calculated their \vz\ and \vb\ velocities.
The relationship between \vz\ and \vb\ is close to 1:1, with \vz\ greater than
\vb\ by around 4.5 kms$^{-1}$ at $b=5$, due to the Sun's own motion in the
Galaxy.
We subtracted this offset and examined the residuals of the \vz\ -- \vb\
relationship to investigate the variance as a function of Galactic latitude
(shown in figure \ref{fig:vb_vz}).
We found that \vb\ is drawn from a heavy-tailed distribution, centered on \vz,
with standard deviation increasing with $b$ (see figure \ref{fig:vb_vz}).
The standard deviation of \vz-\vb\ was around 3kms$^{-1}$ at $b \sim 5^\circ
$, 4kms$^{-1}$ at 10$^\circ$, 6kms$^{-1}$ at 15$^\circ$, and 9kms$^{-1}$ at
20$^\circ$.
This demonstrates that using \vb\ instead of \vz\ for stars in the \kepler\
field will introduce an additional velocity scatter, inflating \sigmavb\
relative to \sigmavz.
This additional velocity scatter will be greatest for stars at the highest
Galactic latitudes.

Since we are concerned with velocity {\it dispersions}, rather than velocities
themselves, we also compared \sigmavb\ and \sigmavz\ as a function of
temperature for stars downloaded from the GUMS simulation.
For stars at galactic latitudes of 15\degrees\ or less, \sigmavb\ was
consistent with $\sigma_{v{\bf z}}$, within uncertainties, however, at higher
latitudes the two quantities became significantly different.
For this reason we proceeded by only including stars with galactic latitudes
less than 15\degrees\ in our analysis.
Although we find that the transformation between \vz\ and \vb\ does not {\it
strongly} affect our results, we cannot rule out the possibility that it
introduces systematic biases into the velocity dispersions we present here.
In \gaia\ DR3, RVs will be available for most stars in this sample, providing
an opportunity to validate (or correct) the results presented here, and to
work in action-space, rather than velocity-space.

Because of the noisy relationship between \vb\ and \vz\, in this paper we do
not attempt to convert velocity dispersion (\sigmavb) into an age via an
age-velocity dispersion relation (AVR) \citep[\eg][]{holmberg2009}.
Although we find that \sigmavb\ can be used to rank populations of stars by
age, a more careful analysis that includes formal modeling of the \vb\ -- \vz\
relationship will be needed to calculate absolute ages.

\racomment{
\section{Appendix B: Photometric temperatures}

For this work, we used the precise \textit{Gaia} DR2 photometric color,
$G_{\rm BP} - G_{\rm RP}$, to estimate \teff\ for the Kepler rotators.
To calibrate this relation, Curtis \etal\ (2020, submitted) combined effective
temperature measurements for nearby, unreddened field stars in benchmark
samples, including FGK stars characterized with high-resolution optical
spectroscopy \citep{brewer2016}, M dwarfs characterized with low-resolution
optical and near-infrared spectroscopy \citep{mann2015}, and K and M dwarfs
characterized with interferometry and bolometric flux analyses
\citep{boyajian2012}.
This empirical color--temperature relation is valid over the color range $0.55
< (G_{\rm BP} - G_{\rm RP})_0 < 3.20$, corresponding to $3070 < T_{\rm eff} <
6470$~K.
The dispersion about the relation implies a high precision of 50~K.
These benchmark data enable us to accurately estimate \teff\ for cool dwarfs
\citep[\eg][]{rabus2019}, and allows us to correct for interstellar reddening
at all temperatures\footnote{The color--temperature relation is described in
detail in the Appendix of, and the formula is provided in Table 4 of, Curtis
\etal\ (2020, submitted).}.
The equation we used to calculate photometric temperatures from Gaia \gcolor\
color is a seventh-order polynomial with coefficients given in table
\ref{tab:coeffs}.
\begin{table}[h!]
  \begin{center}
      \caption{
          Coefficient values for the 7th-order polynomial used to estimate
      \teff\ from \Gaia\ \gcolor\ color, calibrated in Curtis \etal\ (2020, in
      prep).}
    \label{tab:coeffs}
    \begin{tabular}{l|c} 
        (\gcolor\ ) exponent & Coefficient  \\
      \hline
      $0$ & -416.585 \\
      $1$ & 39780.0  \\
      $2$ & -84190.5 \\
      $3$ & 85203.9  \\
      $4$ & -48225.9 \\
      $5$ & 15598.5  \\
      $6$ & -2694.76 \\
      $7$ & 192.865  \\
    \end{tabular}
  \end{center}
\end{table}
}


\ref{tab:data}.
\begin{table}[h!]
  \begin{center}
      \caption{
The data used for this project (the full table is available online).
      Masses were obtained from \citet{berger2020}, rotation periods from
      \mct, effective temperatures were calculated from dereddened \gaia\
      photometry, as described in the appendix and Curtis \etal\ (Submitted).
Velocities were calculated using {\tt pyia} and {\tt astropy}
      \citep{astropy2013, astropy2018, price-whelan2018}.
Uncertainties on velocity dispersions were calculated as the standard
      error of the sample standard deviation \citep{rao1973}.
      }
    \label{tab:data}
\begin{tabular}{cccccc}
KIC ID & Mass [M$_\odot$] & T$_\mathrm{eff}$ [K] & P$_\mathrm{rot}$ [days] & v$_b$ [kms$^{-1}$] & $\sigma_\mathrm{vb}$ [kms$^{-1}$] \\
\hline
1164102 & 0.63 & 4053 & 31.5 $\pm$ 0.5 & 13.6 $\pm$ 0.2 & 21.2 $\pm$ 0.4 \\
1292688 & 0.52 & 3752 & 42.7 $\pm$ 2.1 & -14.2 $\pm$ 0.2 & 21.9 $\pm$ 0.7 \\
1297303 & 0.67 & 4318 & 27.3 $\pm$ 0.2 & 16.7 $\pm$ 0.3 & 16.7 $\pm$ 0.2 \\
1429921 & 0.65 & 4258 & 23.1 $\pm$ 0.1 & -7.0 $\pm$ 0.2 & 14.8 $\pm$ 0.2 \\
1430349 & 0.65 & 4368 & 34.7 $\pm$ 0.8 & 12.2 $\pm$ 0.2 & 19.7 $\pm$ 0.4 \\
1430893 & 0.61 & 3985 & 17.0 $\pm$ 0.0 & 0.4 $\pm$ 0.1 & 9.2 $\pm$ 0.1 \\
1431116 & 0.72 & 4415 & 38.8 $\pm$ 1.0 & -16.5 $\pm$ 0.1 & 19.7 $\pm$ 0.4 \\
1432745 & 0.7 & 4413 & 22.2 $\pm$ 0.2 & -2.1 $\pm$ 0.0 & 13.4 $\pm$ 0.2 \\
1435229 & 0.57 & 4080 & 23.5 $\pm$ 0.0 & -8.8 $\pm$ 1.0 & 13.2 $\pm$ 0.2 \\
1569682 & 0.57 & 3909 & 17.4 $\pm$ 0.2 & 4.4 $\pm$ 0.1 & 9.2 $\pm$ 0.1 \\
\end{tabular}
\end{center}
\tablecomments{Table 2 is published in its entirety in the machine-readable format.
      A portion is shown here for guidance regarding its form and content.}
\end{table}

\bibliography{ms63}{}
\bibliographystyle{aasjournal}



\end{document}